# THGEM operation in Ne and Ne/CH₄

**M. Cortesi**[a*]**, V. Peskov**[afe]**, G. Bartesaghi**[ab]**, J. Miyamoto**[a]**, S. Cohen**[a]**, R. Chechik**[a]**, J.M. Maia**[cd]**, J.M.F. dos Santos**[c]**, G. Gambarini**[b]**,V. Dangendorf**[g] **and A. Breskin**[a]

[a] *Department of Particle Physics, Weizmann Institute of Science, 76100 Rehovot, Israel*
[b] *Department of Physics of the University and INFN Milan, I-20133 Milan, Italy*
[c] *Instrumentation Center, Department of Physics, University of Coimbra, 3004-516 Coimbra, Portugal*
[d] *Department of Physics, University of Beira Interior, 6201-001 Covilhã, Portugal*
[e] *CERN, 1206 Geneva, Switzerland*
[f] *Ecole Nationale Superieure des Mines de St. Etienne, France*
[g] *Physikalisch-Technische Bundesanstalt (PTB), 38116 Braunschweig, Germany*

*E-mail*: Marco.Cortesi@weizmann.ac.il

ABSTRACT: The operation of Thick Gaseous Electron Multipliers (THGEM) in Ne and Ne/CH₄ mixtures, features high multiplication factors at relatively low operation potentials, in both single- and double-THGEM configurations. We present some systematic data measured with UV-photons and soft x-rays, in various Ne mixtures. It includes gain dependence on hole diameter and gas purity, photoelectron extraction efficiency from CsI photocathodes into the gas, long-term gain stability and pulse rise-time. Position resolution of a 100x100 mm² X-ray imaging detector is presented. Possible applications are discussed.

---

* Corresponding author.

# Contents

# 1. Introduction

THick Gaseous Electron Multipliers (THGEM) are building blocks of gaseous radiation detectors [1]. Avalanche multiplication of radiation-deposited electrons occurs within 0.2-0.7 mm diameter holes, drilled through 0.4-0.8 mm thick double-face copper-clad insulator plates; etched rims ($\leq 0.12$ mm wide) around the holes reduce discharge probability. The rim is the main distinction from the "optimized GEM" [2] and the "LEM" [3] THGEM electrodes can be mass-produced by standard printed-circuit board techniques (using photolithographic masks); as robust and mechanically self-supporting elements they are suitable for large-area applications. High charge gains were demonstrated with single- and double-THGEM elements in a variety of gases [4], including noble gases and their mixtures [5]. For a review and more complete bibliography on THGEM operation and properties (e.g. gain, energy and time resolutions, stability etc.), the reader is referred to ref. [6]. Their operation in cryogenic conditions [7], including two-phase mode in Ar, was recently demonstrated [8]. Potential applications are in large-volume detectors for rare events, large-area UV-photon imaging detectors (e.g. coated with CsI photocathodes) for RICH [9-11], sensing elements in digital calorimetry [12], moderate (sub-millimetres) resolution particle-tracking as well as X-ray and neutron imaging detectors.

This work summarizes the results of recent investigations of THGEM detectors operated with Ne and Ne/$CH_4$ mixtures at atmospheric pressure. The reason for using Ne and its mixtures is their much lower operation potentials, compared with "traditional" Ar-mixtures. This is due to the relatively higher value of the first Townsend coefficient in Ne, which characterizes the avalanche process as a function of the electric field. Accordingly, the onset of charge multiplication by electron-impact ionization in Ne takes place at lower electric fields compared to heavier noble gases.

The low operation voltage has some substantial advantages: it significantly reduces field emission due to mechanical-drilling defects and discharge probability induced by highly ionizing events; it moderates the spark-energy (in case of a discharge), which is proportional to the square of the voltage difference applied across the electron multiplier; it induces lower charging-up effects to the insulator and, finally, it should permit reaching high-gain operation at high pressures, which could be of interest in some applications.

Ne is being considered as filling gas in detectors for low radioactive background physics experiments, mainly because it does not require expensive purification systems like Xenon for example. In addition, Ne does not contain long-living radioactive isotopes, whose background signals interfere with the signals from rare physics events. For example, high-pressure Ne gas (200 bar) is being considered as target medium for dark matter detection and coherently scattered neutrino studies [13]. Supercritical Ne phase is exploited for a TPC designed to detect low-energy solar neutrinos; it relies on the formation of slowly-drifting electron bubbles (10 cm/sec) towards a read-out plane, undergoing low diffusion [14].

Large-volume Ne-filled detectors (including two-phase devices) with THGEM readout elements of charges or/and scintillation light, could be employed for the detection of rare physics events. The electrodes could be made of low-radioactivity materials, e.g. CIRLEX (a polyimide) [15].

Though the properties of THGEMs with various gas-filling have been studied in great detail (see [6] and references therein), Ne-filled THGEMs required further investigations. The main points of concern being secondary effects, due to ions and to the energetic (15.5 eV) VUV avalanche-induced photons; the relatively large electron diffusion, affecting their collection into the THGEM holes and the low ionization threshold as a function of voltage, initiating avalanche multiplication out of the THGEM holes. The fact that the avalanche may no longer be confined within the holes can trigger and enhance secondary effects, mediated by ions and photons; these affect the achievable gain and other properties of the detector like energy-, time- and position-resolutions. Indeed, in tests of resistive THGEMs [16] in Ne, secondary effects were observed at high gas gains. The photon-induced secondary effects can be reduced by "quenching" the energetic photons, e.g. with $N_2$, $CH_4$ and other molecular gases. Among Ne mixtures previously used in other detector types are: Ne/dimethylether (DME) [17] and Ne/$CF_4$ [18] and Ne/$H_2$ [19].

In this work we studied the properties of THGEM operation in Ne and Ne-$CH_4$ mixtures, with soft X-rays and UV-photons; the latter were investigated using a CsI photocathode (PC) deposited on the first THGEM cathode surface. The properties investigated were: gain vs voltage, maximum achievable gain, role of impurities, role of hole-geometry, electron transport in multi-electrode THGEM assemblies, photoelectron extraction from the PC, energy resolution, long-term gain-stability and imaging properties.

## 2. Experimental setup and methodology

The basic setup for most studies is shown in Fig. 1. It consisted of two THGEM electrodes (manufactured by "Print Electronics", Israel) in cascade; they could be operated either in single-THGEM or in double-THGEM mode. Electrodes of various geometrical configurations [4] (variable parameters are: thickness "t", hole diameter "d", hole spacing "a" and rim-size "h"), and with an active area of 30x30 or 100x100 mm$^2$, were investigated. The setup was mounted in a stainless steel vacuum vessel, possibly equipped with two windows: a UV transparent window made of Suprasil and a thin (25 micron thick) polymer foil for low energy X-rays transmission, both of 1 cm in diameter. This permitted alternate investigations of a given detector setup with either UV photons from a continuously emitting Ar(Hg) lamp or alternatively with soft X-rays, without opening the vessel. The vessel was evacuated to $10^{-5}$-$10^{-6}$ mbar with a turbo-molecular pump, prior to gas introduction. The gas was flushing continuously through the vessels, in flow mode.

As presented and discussed in section 3.6, our investigations indicated that the electron avalanche processes in Ne and Ne mixtures (and consequently gain-curve shape, operation voltage and maximum reachable effective gain), with no quenching admixtures, are extremely sensitive to the amount of impurities present in the gas. Similar observations were made in Ar-operated THGEMs [5]. The main sources of impurities are residual molecules within the vessel, contaminants present in the detector components and in the gas system (mainly moisture and hydrocarbons), air leaks etc. [19]. In general, it is well known that the gas-impurities exchange charge with ionized Ne (see for example [20]) and may at their high concentration also absorb avalanche VUV photons. For instance, $N_2$ at some conditions may act as a wavelength shifter, from VUV-photons produced by Ne disexcitation (wavelength $\geq$ 74 nm) to the visible range (300 to 450 nm) [21]. At longer wave length photons can extract photoelectrons from the CsI photocathode or from other metal-coated detector components, but with very low quantum efficiency.

At low concentration of impurities (<10-50 Torr) however the gas become transparent for the UV emission of the avalanches and only the charge exchange mechanism and the suppression of the Ne excimer emission contribute to the quenching mechanism

We observed that THGEMs operating in Ne, with some small $N_2$ impurities (or admixture) provide stable high-gain operation.

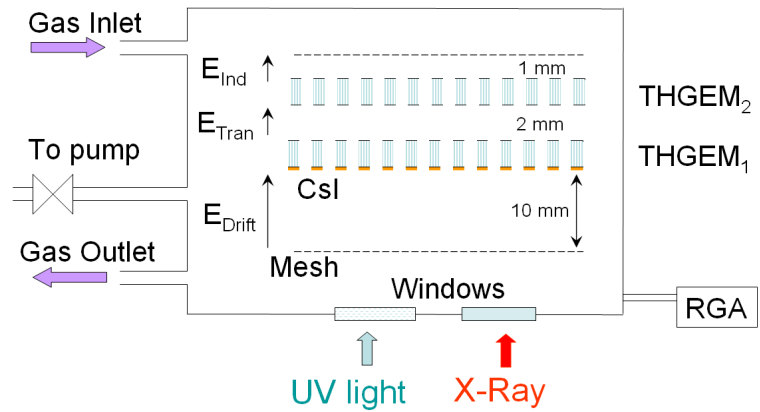

**Figure 1.** Schematic drawing of the THGEM detector investigations set-up (RGA = residual gas analyzer for monitoring the gas composition). The different gap dimensions are given as example and varied according to the experiment.

In gas-flow mode, the impurities content is gradually reduced and stabilized by continuous flow. The stabilization time and residual contaminants level depend on the flow rate and on the gas-system components and materials. In our systems, it was found that gas flushing for at least 24 hours, at a constant rate of 0.2-0.4 cc/m, stabilizes the impurity composition in Ne and Ne mixtures within the detector vessel, reaching stationary operation conditions. In some setups the gas purity could be monitored online with a residual gas analyzer (RGA); in stationary condition, the level of contaminants was of the order of 0.5%-1% (therefore, note that, whenever we refer to Ne in the text below, it is actually Ne + contaminants, at a level around 1%). In long-term measurements, care was taken to monitor pressure and temperature; in this case the data were corrected for variations or these parameters.

The measurements were carried out both in continuous mode (current-mode or DC mode, recording currents from various electrodes, and in pulse-counting mode – measuring pulses with charge-sensitive preamplifiers (ORTEC 124), followed by shaping amplifiers (ORTEC 572A) and further recorded by multi-channel analyzers (Amptek MCA8000A). The high voltage was supplied to individual THGEM electrodes through 15 MΩ serial resistors to limit possible discharge currents. Current limits of 50 nA were usually set on the power supplies (CAEN, model N471A).

The photoelectron extraction efficiency ($\varepsilon_{Ex}$), from a CsI PC deposited on top of the first THGEM cathode surface, and the combined extraction and electron collection efficiency ($\varepsilon_{Ex}$ $\varepsilon_{Coll}$) into the THGEM holes (under THGEM-gain = 1) were measured in Ne, Ne/CH$_4$ and Ar/CH$_4$ mixtures by means of the set-up shown in Fig. 2. These experiments were carried out with two 30x30 mm$^2$ cascaded THGEMs (FR4 material t = 0.8 mm, d = 0.5 mm, h = 0.1 mm and a = 1 mm) placed 3 mm apart, between two stainless-steel (81% transparency) meshes; the mesh-to-THGEM distances (drift gaps) were of 1 cm. Both multipliers were illuminated in turn by a UV Ar(Hg) lamp through the corresponding Suprasil UV-windows; one of the two multipliers (THGEM$_b$ in Fig. 2) was coated with a CsI photocathode layer (0.35 μm thick), while the other one (THGEM$_a$ in Fig. 2) had a bare Cu surface. Besides measurements of photoelectron extraction efficiency, this procedure allowed for comparing photocurrents emitted from the CsI-coated THGEM and the bare (Cu) THGEM; possible contribution from Ne scintillation (excimer radiation) in high electric field could be investigated.

The photocurrents emitted from both THGEM top electrodes, were measured with picoampermeters as function of the applied fields in the drift gaps – the extraction fields. The photoelectron extraction efficiency in gas was calculated in two steps: first we measured the photocurrent from the top THGEM electrodes as a function of the applied voltage $V_{Drift}$ between the drift mesh and the top electrode in vacuum (see the setup in Fig 2a). Typically this current sharply increased with the applied voltage and reached a clear plateau $I_{vac}(V)$ = constant, at $V_{Drift}$ > 100V. Then the same measurements were repeated in gas, providing $I_{gas}$ (V). The extraction efficiency was defined as $\varepsilon_{Ex}(V_{Drift}) = I_{gas}(V_{Drift})/I_{vac}$ (Fig. 2a).

The combined electron extraction and electron collection efficiency into the holes were derived from the current measurements at the bottom of the THGEM, after focusing the photoelectrons into the holes (Fig. 2b). In mixtures of Ne with CH$_4$ this current increased with the voltage applied across the THGEM ($V_{THGEM}$) reached a plateau current ($I_{THGEM}$) and started to rise sharply at higher fields due to the onset of gas multiplication in the holes. The combined extraction and electron-collection efficiency into the THGEM holes was defined as $\varepsilon_{Ex} \cdot \varepsilon_{Coll} = I_{THGEM}/I_{vac}$.

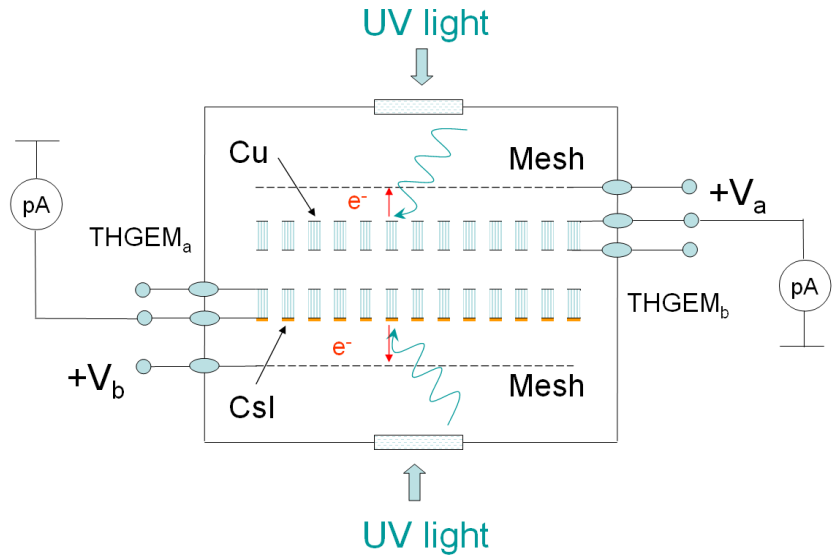

**Figure 2a.** Schematic drawing of the THGEM detectors set-up for extraction efficiency measurements.

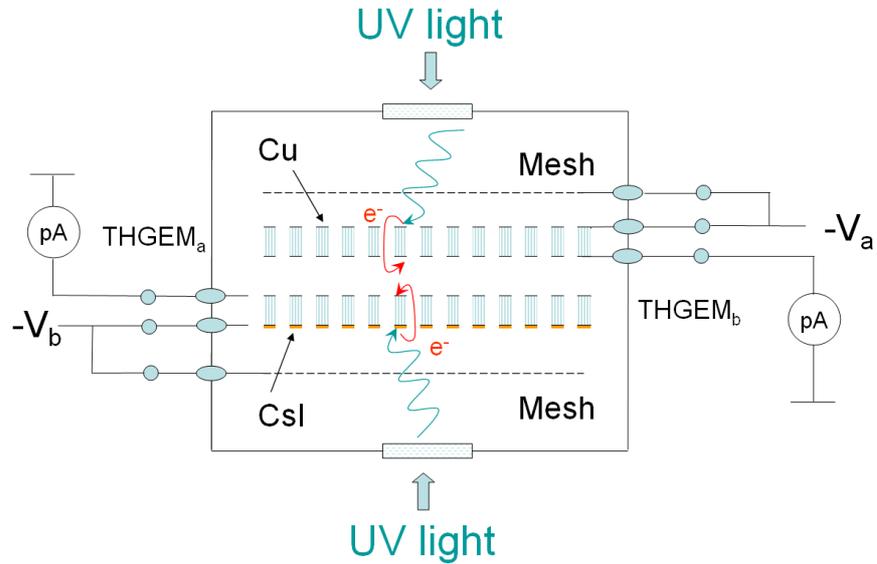

**Figure 2b.** Schematic drawing of the THGEM detectors set-up for extraction combined with collection efficiency measurements.

The imaging performance of a THGEM-based detector prototype was measured using the experimental configuration shown in Fig. 3. The detector comprised of two 100x100 mm² THGEMs in cascade, coupled to a dedicated resistive anode read-out electrode [22], equipped with processing electronics for 2D position encoding. The data acquisition hardware (DAQ) was based on an 8-channel Time-to-Digital Converter card (RoentDek TDC8PC) while data acquisition and analysis program were performed with RoentDek CoboldPC program package [23].

The imaging THGEM-based detector was located within a stainless steel vessel, having a 25 μm thick Mylar window, 20 cm in diameter. It was irradiated with continuous Bremsstrahlung spectrum originating form 5-9 keV electrons hitting a Cu target. The X-rays conversion gap above the first THGEM multiplier was kept at 10 mm; the transfer and induction gaps were 2 and 1 mm wide, respectively. The detector was vacuum pumped (~ 0.1 mbar) and then flushed with Ne or Ne/5%CH$_4$ at atmospheric pressure.

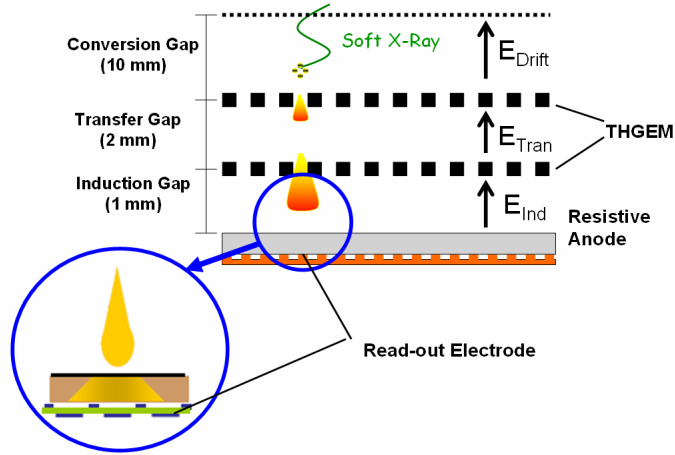

**Figure 3.** The THGEM-based imaging detector set-up.

## 3. Results

### 3.1 Gain

Fig. 4 depicts single-THGEM effective-gain curves in Ne, Ne/CH$_4$ mixtures and Ar/5%CH$_4$ at atmospheric pressure, obtained under the same experimental conditions (under gas-flow and in continuous current mode). The measurements were performed with a THGEM electrode (t = 0.8 mm, d = 0.6 mm, a = 1 mm, h = 0.1 mm) equipped with a reflective CsI photocathode and illuminated with UV light. Also in Fig. 4, the gain limits are indicated, which were obtained in each condition while irradiating the same detector (of Fig. 1) with Cu 9 keV X-rays.

The highest (defined by the onset of discharges) effective-gain value, of a few times 10$^6$ with single photoelectrons, was reached in Ne at about five-fold lower operation voltage compared to Ar/5%CH$_4$. The addition of a few percent of methane to the Ne reduced the corresponding first Townsend coefficient of the gas mixture. Consequently, the effective-gain curve shifted towards higher operation voltages with increasing methane fraction.

It is interesting to notice that in Ne and Ne/CH$_4$ mixtures the maximum reachable gain with X-rays irradiation (~ 250 primary electrons) was only 4-fold lower compared to that with single-photoelectrons, while in Ar/5%CH$_4$ it was two orders of magnitude lower. The higher breakdown limits with X-rays in Ne-mixtures can be attributed to the relatively lower charge avalanche density in Ne-mixtures, due to a longer range of the photoelectron and higher electron diffusion, resulting in primary charge spread and thus avalanche spread over a few holes; this could relax the constraint imposed by the known Reather limit [24]. The effect is particularly prominent in the case of Ne/23%CH$_4$ and Ar/5%CH$_4$, having almost identical gain curves but a 10-fold lower discharge limit for X-rays in the latter.

The effective-gain curves obtained with two 100×100 mm$^2$ THGEMs in cascade (t = 0.4 mm, d = 0.5 mm, a = 1 mm and h = 0.1 mm), operating in various Ne-methane mixtures is depicted in Fig. 5. The detector was irradiated with continuous Bremsstrahlung spectrum originating form ca. 9 keV electrons impinging on a Cu target. The maximum achievable effective gain monotonically dropped and the operation voltages increased with increasing CH$_4$ fraction. It is further noted that the maximal gain increases with the THGEM thickness; e.g.

with ca. 9-keV X-rays in this double THGEM it is typically only 5-10 higher than with a single, twice thicker, THGEM detector (of Fig. 4).

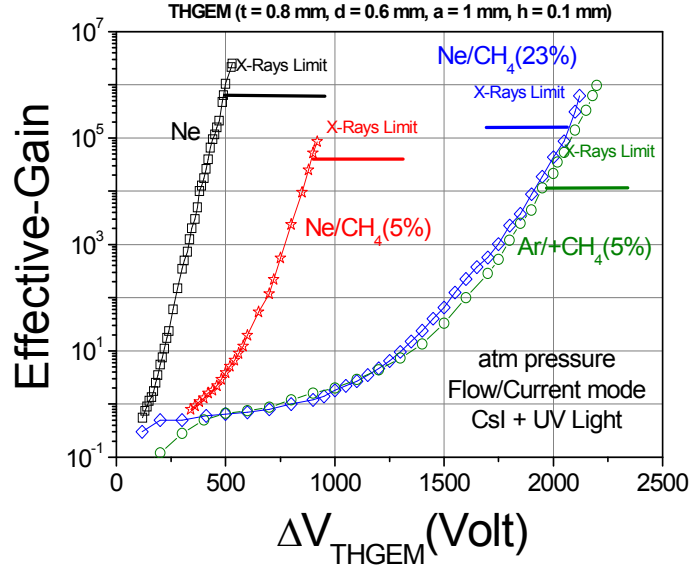

**Figure 4.** Effective-gain curves obtained in the single-THGEM detector of Fig. 1, with single UV photons (CsI photocathode; $E_{drift} = 0$) in Ne, Ne/CH$_4$ mixtures and in Ar/CH$_4$ (95:5). The maximum effective gain reached in each gas, in the same detector, with 9 keV (Cu-K$_\alpha$) X-rays is also shown ($E_{drift} = 0.2$ kV/cm). THGEM geometry: t = 0.8 mm, d = 0.6 mm, a = 1 mm, h = 0.1 mm.

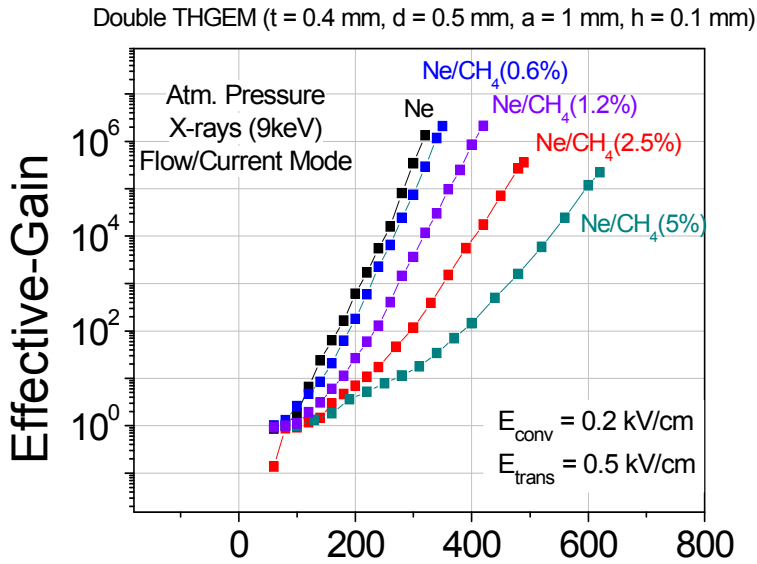

**Figure 5.** Effective-gain curves in the double-THGEM imaging detector of Fig. 3; data measured in Ne and Ne-CH$_4$ mixtures, in gas flow-mode, with 9 keV X-rays. THGEM geometry: t = 0.4 mm, d = 0.5 mm, a = 1 mm, h = 0.1 mm.

Figure 6 compares the effective gains measured in single- and double-THGEM (both have the same thickness t = 0.4 mm) in Ne gas (flow-mode), irradiated with UV light and with ca. 9 keV X-rays. It is evident that a double-electrode cascade has close to 100-fold higher total gain, at almost two-fold lower THGEM operating voltage, clearly demonstrating the advantage of a cascade configuration. Similarly to Fig. 4, 10-fold lower maximal gains were reached with X-rays compared with single-photoelectrons, in both single- and double-THGEM configurations (respective gains of ~ $10^4$ and $10^6$).

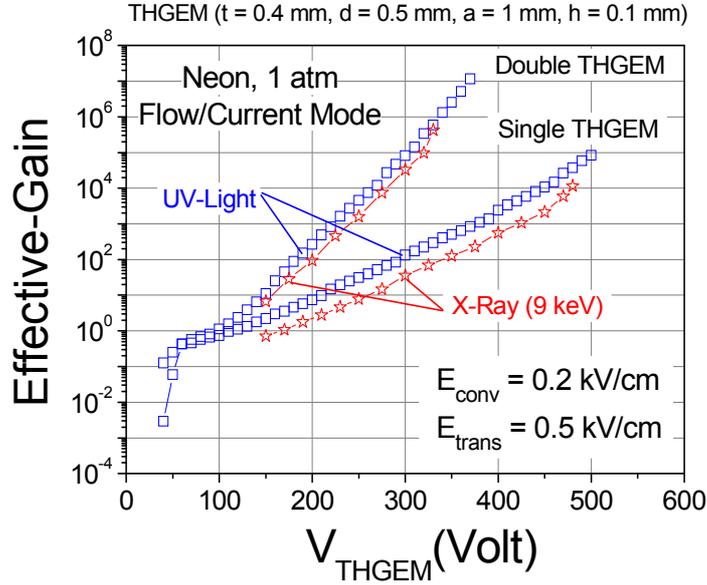

**Figure 6.** Effective-gain curves in Ne of single- and double-THGEM detectors, measured with single UV photons and with soft X-rays. THGEM geometry: t = 0.4 mm, d = 0.5 mm, a = 1 mm and h = 0.1 mm.

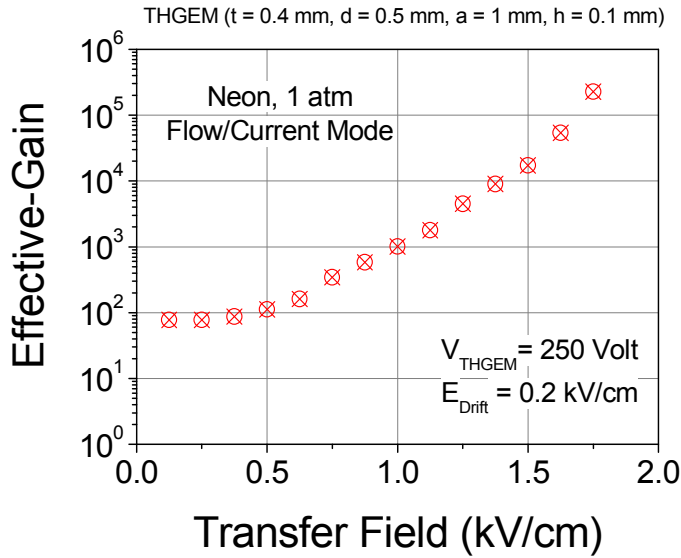

**Figure 7.** Effective-gain variation as function of the transfer field between two THGEMs, with their operation potentials kept at 250 Volt and with $E_{Drift}$ = 0.2 kV/cm (Detector of Fig. 1). THGEM geometry: t = 0.4 mm, d = 0.5 mm, a = 1 mm, h = 0.1 mm.

In Ne and Ne-mixtures charge multiplication occurs already at very low fields, and therefore it might take place also within the drift, transfer and induction gaps if these exceed a certain limit (Fig. 1). In particular, it might take place at the hole vicinity, within the dipole field extending outside the hole into the gaps. Fig. 7 shows the effective gain measured in a double-THGEM as function of the transfer field between the two elements: the voltage difference across each multiplier was kept at a value of 250 Volt (gain $\sim 10^2$; Fig. 6). At high transfer fields (above the multiplication threshold of $\sim 0.4$ kV/cm) the total detector's effective gain ($G_{Total}$) may be expressed as the product of:

$$G_{Total} = G_{THGEM} \, G_{Tran} \tag{1}$$

where $G_{THGEM}$ is due to the THGEM multiplication while $G_{Tran}$ takes into account the parallel-plate like multiplication occurring within the transfer gap.

Figure 8 presents the energy resolution of 5.9 keV X-rays, as function of the drift field, for single-THGEM detectors operated in Ne and Ne-methane mixtures (2.5% and 5% of $CH_4$) at an effective gain of $\sim 10^4$. Energy resolutions of $\sim 30\%$ FWHM were measured in all gases, with a single-THGEM (t = 0.8 mm, d = 0.5 mm, a = 1 mm, h = 0.1 mm), over a wide range of drift fields; it indicates at a good and constant overall collection efficiency of the ionization electrons into the THGEM holes Slightly better energy resolutions ($\sim 25\%$ FWHM) were obtained with drift field = 0. In this condition only the X-rays converted at the holes' vicinity were detected; they induced electron clouds that were most efficiently collected by the dipole field and thus resulted in better energy resolution. The steep deterioration of the energy resolution, at 0.35 kV/cm for Ne and 0.7 kV/cm for Ne/2.5%$CH_4$, is due to the onset of multiplication in the drift gap.

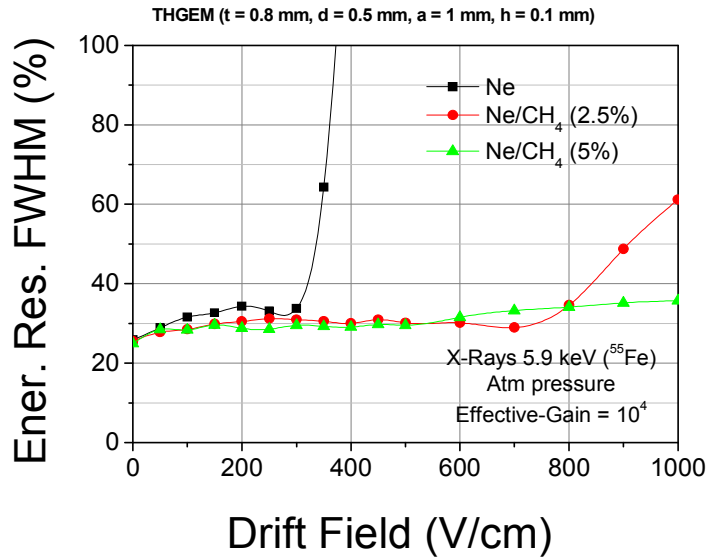

**Figure 8.** Energy resolution of 5.9 keV X-rays, measured in a single-THGEM at an effective gain of $\sim 10^4$, as function of the drift field, in Ne and Ne/$CH_4$ mixtures. THGEM geometry: t = 0.8 mm, d = 0.5 mm, a = 1 mm, h = 0.2 mm.

We may conclude that in order to maintain best performance when using Ne and Ne mixtures, care should be taken in the choice of drift and transfer fields, such that the multiplication is not occurring in the gaps.

### 3.2 Pulse-shape

Signal rise-time in Ne and Ne-$CH_4$ mixtures ($CH_4$ at 1.2%, 3% and 5% concentrations) was measured in double-THGEM configuration, irradiating the detector with ca. 9 keV Bremsstrahlung X-rays. The detector signals were processed by a low-noise fast preamplifier (VV44, MPI Heidelberg) and a shaping amplifier (Ortec 570). Fig. 9 shows pulses recorded in Ne and Ne/5%$CH_4$ at a gain of $\sim 10^5$; the respective rise-times were $\sim 70$ ns and $\sim 30$ ns. The decrease of rise-time with increasing $CH_4$ concentration in the mixture (Fig. 9c) is due to the increase in electron drift velocity.

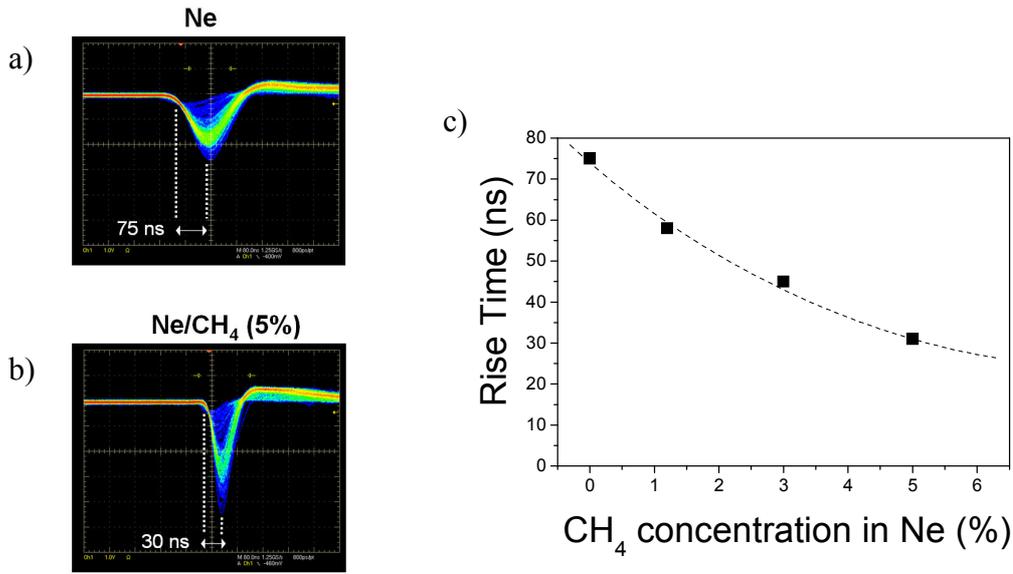

**Figure 9.** Signals from the double-THGEM of Fig. 3 in Ne (a), and in Ne/$CH_4$ (5%) (b), recorded with ca. 9 keV X-rays. The signal's rise-time as function of $CH_4$ percentage is shown in c).

### 3.3 Long-term gain stability

Long-term gain stability is an important and relevant issue in multipliers based on insulating substrates, or having insulator surfaces exposed to the high electric fields. It is well known [25] that hole-type gas multiplication structures, made of holes etched or drilled in insulator materials, suffer from initial gain variations after applying high voltage to the electrodes (charging-up effect). The gain stabilization-time depends on the polarization of the dielectric substrate and on the accumulation of charges on the dielectric surfaces exposed to the avalanche processes. While the former depends on the dielectric material (e.g. surface resistivity) and on the applied voltage, the latter depends on the detector's gain and on the counting rate per unit area.

Measurements of long-term gain stability are tedious and often not reproducible due to difficulties in precisely reproducing the experimental conditions: gas purity, impurities from gas pipe system, substrate material, and content of moisture and residual chemicals in the substrates, all affecting surface resistivity. An attempt to perform a systematic study of the gain stability was carried out in a single-THGEM with a reflective CsI photocathode, irradiated with UV light; the detector was operating in various gas mixtures, recording the total anode current evolution in time, shown in Fig. 10. The measurements were performed with a fixed effective-gain value, of $\sim 10^4$, and a photoelectron flux of the order of $10^4$ $s^{-1}$ $mm^{-2}$. The gas was introduced into the vessel after pumping to a vacuum of $10^{-5}$-$10^{-6}$ mbar: the data were recorded after 24 hours of gas flow.

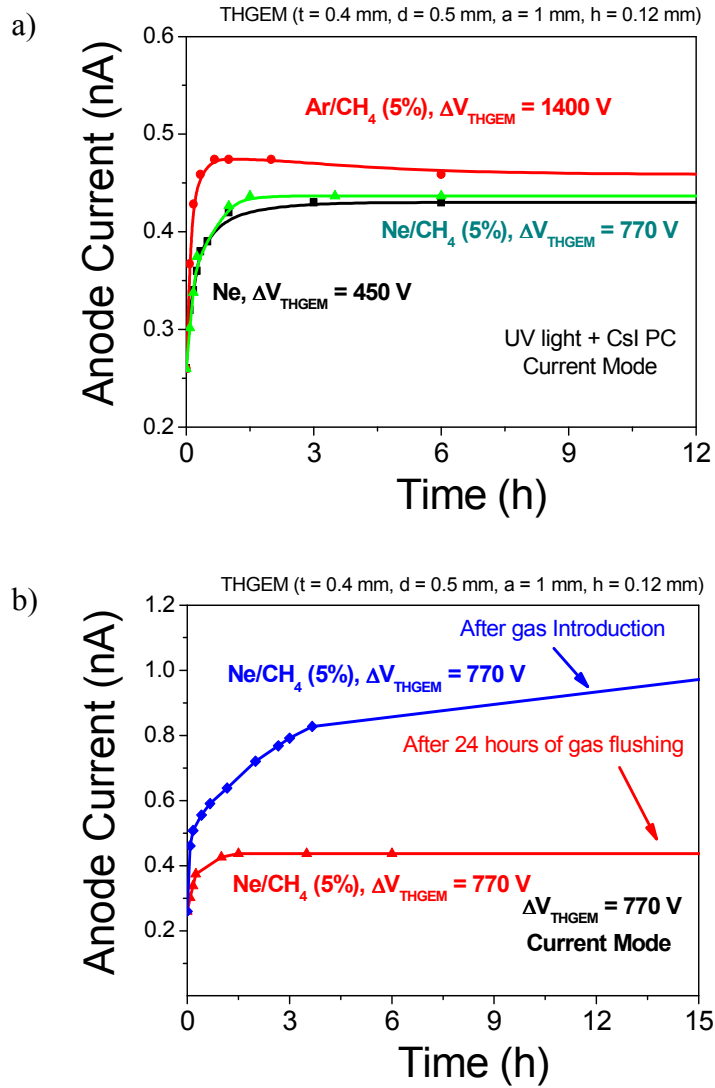

**Figure 10.** Gain stability under UV-photon irradiation of a single-THGEM coated with a CsI photocathode, measured in a): Ne, Ne/5%CH$_4$ and Ar/5%CH$_4$ after 24 hours of gas flow prior to measurements; b) Ne/5%CH$_4$ after 24 hours of gas flow compared to measurements taken immediately after gas introduction. Effective gain: ~ $10^4$; counting-rate of ~ $10^4$ photoelectrons per second and per mm$^2$.

As shown in Fig. 10a, similar stabilization-times were reached in Ne and in Ne/CH$_4$ ($\approx$ 1-2 hours); in Ar/CH$_4$, due to the higher operation voltage, we observed an "overshoot" due to the substrate's polarization. Fig. 10b shows that during ~ 6 hours, a 4-fold gain variation was recorded when the measurement started immediately after pumping and gas introduction; this compares to a 2-fold increase if the measurements started after prolonged flow of gas prior to the measurements. The difference is probably related to changes in the content of gas impurities during this prolonged gas flow, which affect the surface resistivity of the THGEMs.

### 3.4 Maximum achievable gain versus irradiation rate in Ne

The relation between maximum achievable gain and the irradiation rate in micro-pattern gas detector was reported and discussed in detail by several authors [26-28]. While at low rates, the Reather condition (streamer mechanism) and ion/photon feedback loops may be the main causes for breakdowns, at high irradiation rates breakdown may appear due to the "cathode excitation". The cathode excitation mechanism may be explained by a formation of temporal double ion layer on conductive surfaces; it reduces the work function and may cause electrons' emission in forms of jets or bursts which may initiate discharges [29]. The probability for transition from proportional mode to discharge, as a consequence of the Reather limit or cathode excitation mechanism, strongly depends on the radiation flux density as well as on the composition of the gas-mixture. As a general trend, in mixtures with smaller quencher content the avalanche spreads over larger volumes; it results in reduced charge density, pushing the sparking-limit to higher irradiation flux.

The maximum achievable effective gain was measured as function of the irradiation rate of a single -THGEM ($30x30$ mm$^2$) detector with a reflective CsI photocathode (Fig. 1), for a range of photoelectron flux density spanning from $10^{-2}$ up to $10^5$ s$^{-1}$ mm$^{-2}$. The detector was operating in Ne at atmospheric pressure. As shown in Fig. 11, the maximum achievable effective gain systematically drops as the irradiation rate increases.

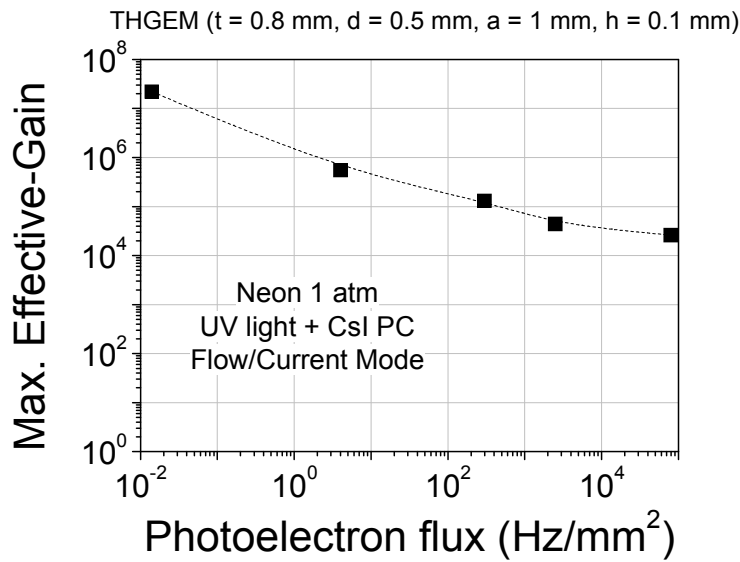

**Figure 11.** Maximum achievable effective gain versus radiation flux density (photoelectron flux) for a single-THGEM (of Fig. 1) operating in Ne at atmospheric pressure.

Due to the large electron diffusion coefficient in pure Ne and the advantages of hole-type structures in moderating secondary effects, gains in excess of $10^4$ were reached even at photoelectron flux density of $10^5$ s$^{-1}$ mm$^{-2}$.

Another phenomenon related to irradiation rate in avalanche detectors is the space-charge effect, due to slowly-moving ions modifying the field at the anode vicinity and reducing the gain at high rates. In micro-pattern electrodes the ion collection is generally accomplished within few microseconds, thus space-charge effects are not apparent below rates of 100kHz/mm$^2$. It was shown earlier [4] that the THGEM's effective gain exhibits a flat response up to high radiation rates, e.g. up to MH/mm$^2$ with single photoelectrons at gain of $10^4$.

### 3.5 THGEM geometry and maximum achievable effective gain

Figure 12a depicts the effective-gain curves obtained with double-THGEMs of different hole diameters; all the other geometrical parameters (thickness t = 0.4 mm, pitch a = 1 mm for d < 1 mm, pitch a = 2 mm for d = 1.2 mm and rim size h = 0.1 mm) were kept constant. All data were recorded under the same experimental conditions (vessel, gas filling and irradiation procedure).

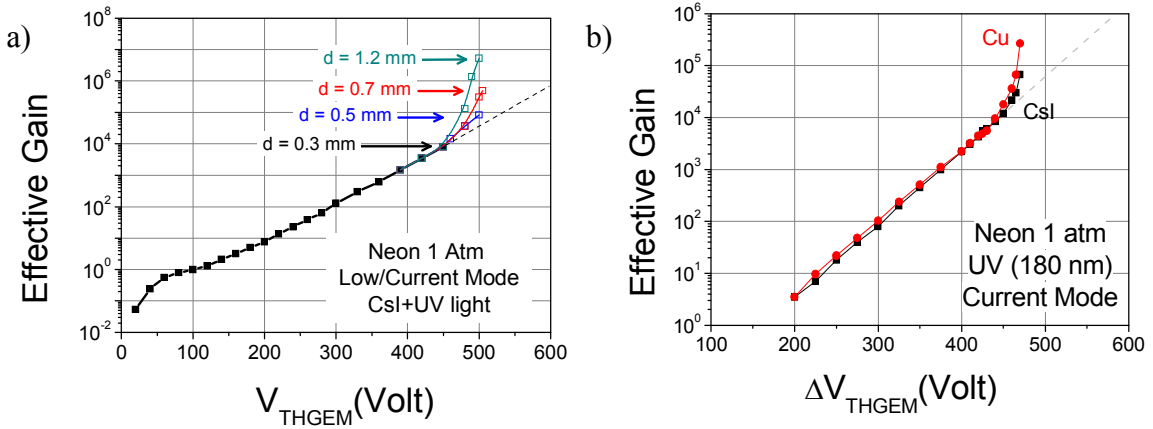

**Figure 12.** a) Single-THGEM gain curves in Ne at atmospheric pressure, measured with electrodes of different holes' diameters. The arrows indicate maximum achievable effective gains. b) Gain-curves measured in Ne with a CsI-coated THGEM and with bare-Cu THGEM at UV flux densities of around 250 s$^{-1}$ mm$^{-2}$. THGEM geometry: t = 0.4 mm, d = 0.3-1.2 mm, a = 1-2 mm and h = 0.12 mm.

These data show that for the similar reachable maximum voltages, the effective-gain values increase with the holes' diameter –diverging from the exponential line above gain-values of ~ 10$^4$. The divergence, indicating appearance of secondary avalanches, increases with increasing hole-diameter, the latter also known to be related to increasing extension of the dipole field out of the holes. We suppose that in Ne, due to the low multiplication threshold (e.g. Fig.7), this also involves the extension of the avalanche outside of the holes, because the sizeable electric field out of the holes is sufficient for inducing multiplication in the gas region above and below the THGEM electrode. In this case the avalanche is no longer confined inside the hole, and secondary effects become important.

In order to find out whether the secondary effects are due to the presence of CsI PC and its interaction with avalanche photons and ions, we carried out a comparative measurement, with CsI-coated THGEM and bare Cu THGEM, under otherwise similar conditions. The gain curves measured in Ne with both THGEMs (both of identical geometry: t = 0.8 mm, d = 0.5 mm, a = 1 mm and h = 0.1 mm) are shown in Fig. 12b. Only ~ 3 times higher gain was recorded with the bare THGEM indicating that the presence of the CsI photocathode did not seriously affect the maximum achievable gain in this THGEM geometry.

The electric field strength and direction within a THGEM cell were computed with Maxwell 3D electric field simulator [30], for typical $\Delta V_{THGEM}$ values of 300-500 V. Fig. 13a depicts the electric field strength calculated along the hole's axis for various $\Delta V_{THGEM}$ values (t = 0.4 mm, d = 0.5 mm, a = 1 mm, h = 0.1 mm); the figure shows the growing values of the field strength extending out of the hole, with the applied voltage, namely the growing probability for secondary effects in Ne. For example, electric field above the avalanche onset threshold, of 0.4 kV/cm, extends out of the 0.5 mm diameter hole by about 0.4 mm, even with $\Delta V_{THGEM} \approx 300$ V (corresponding gain ~ 10$^2$); this in principle should have provoked secondary effects already at very low gain. However, in practice the Ne in our system contained minute amount of impurities and therefore the scintillation yield at low fields was quite low, [21], and photon-mediated secondary effects were relevant and observable only at higher electric-field strengths (see deviation from an exponential behavior in Fig. 12a).

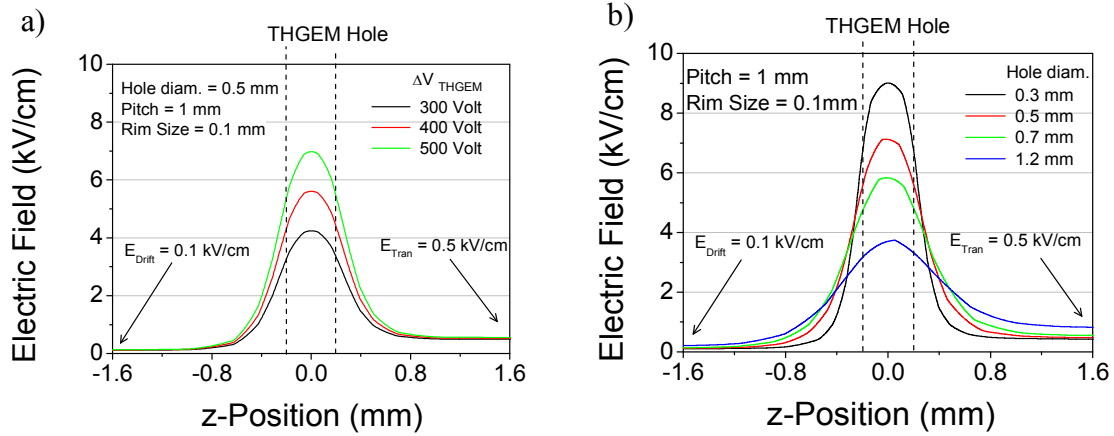

**Figure 13.** a) depicts the electric field calculated on the hole's axis for various voltage differences applied across the THGEM while b) depicts the electric field strength along the hole's axis for different hole's diameters.

Fig. 13b depicts the electric field calculated along the hole axis at constant $\Delta V_{THGEM} \approx 300$ V for various holes' diameters. One observes better field confinement within smaller-diameter holes – namely smaller avalanche extension and smaller probability for secondary effects, in consistency with the data depicted in Fig. 12a. Nevertheless, small amount of photon-quenching admixtures (e.g. $CH_4$) strongly reduce photon-mediated secondary effects, even for large hole diameters.

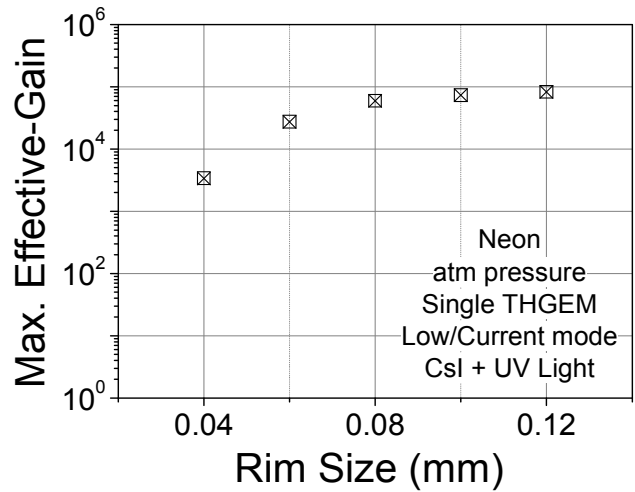

**Figure 14.** Maximum effective gain of a single-THGEM in Ne at 1 atm as function of the hole's rim size. THGEM parameters: t = 0.4 mm, d = 0.5 mm and a = 1 mm.

It was shown earlier [6, 25], with other gas fillings, that the maximum achievable effective gain of THGEM-based detectors, depends on the size of the etched rim around its holes. In this work we investigate this effect in Ne. Fig. 14 depicts the maximum gain reached in a single-THGEM in Ne; it shows that even though the voltages in Ne are considerably lower than in other mixtures studied in [6, 25], the rim size is very important for high achievable gain: electrodes with rim sizes above 0.08 mm provided twenty-fold higher gains compared to holes with small rims (0.04 mm). Indeed, electrodes manufactured with small holes' rims may have higher probability for occasional mechanical defects that would limit the multiplier's performances. It is however worth recalling that there is a trade-off, and better gain stabilities were reached in Ar/30%$CO_2$ with very small (microns) rims, though at low gains [4].

### 3.6 Gas purity and maximum achievable effective gain

Figure 15 shows effective-gain curves versus THGEM voltage, measured in a double-THGEM of Fig. 1 (Parameters are: t = 0.4, d = 0.6, a = 1 and h = 0.1 mm), at 1 bar of Ne. The drift field ($E_{drift}$) was set to 0.1 kV/cm, while both transfer ($E_{trans}$) and induction fields ($E_{ind}$) were set at 0.5 kV/cm. Measurements started when flushing Ne into an air-filled vessel (e.g. high amount of air) and lasted for about 20 hours; the impurities level are specified for $N_2$, $O_2$ and $H_2O$. Note the decrease in operation voltages with the decrease of these contaminants, and the $\sim$ 30-fold reduction in gain with the $N_2/O_2/H_2O$ content decreasing from 15%/3%/2% to 0.1%/0.001%/0.2%. It is expected that the level of other impurities, e.g. Ar and hydrocarbons, varied accordingly.

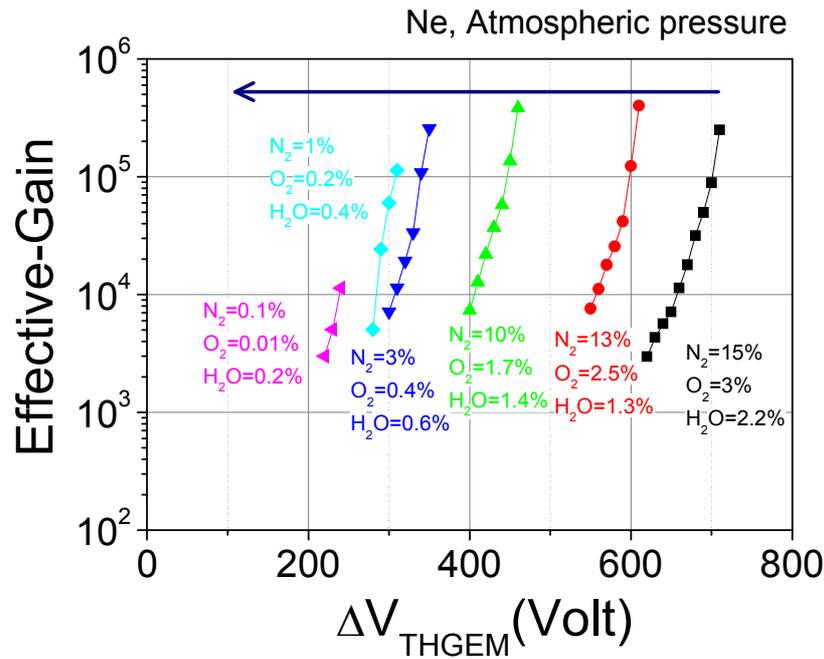

**Figure 15.** Gain curves in a double-THGEM detector, measured in a vessel containing residual air fraction, and being continuously under Ne gas flow; the impurity percentage is indicated for $N_2$, $O_2$ and $H_2O$. The time span for the measurement was $\sim$ 20 hours.

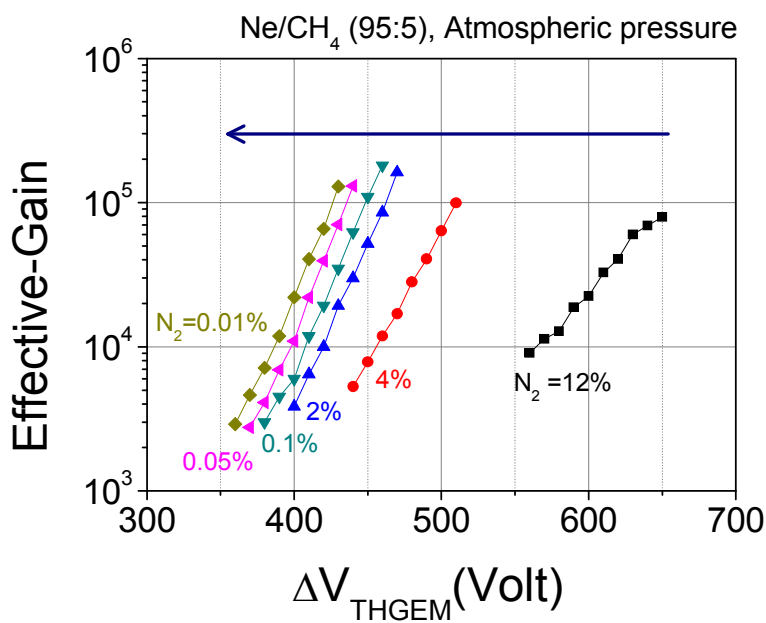

**Figure 16.** Effective-gain in a double-THGEM detector operated with 1 atm. Ne/5%CH$_4$, in a vessel containing residual air fraction and being continuously under gas flow; the overall impurity percentage is represented by N$_2$ percentage in Ne. The time span for the measurement was ~ 20 hours.

The situation is different when a fix amount of one quencher was introduced, e.g. CH$_4$ to Ne; as shown in Fig. 16, the considerable changes in impurities level (e.g. N$_2$-level variation from 12% down to 0.01%) did not significantly affect the maximum effective-gain value; the only effect was the considerable decrease of the operation voltage with decreasing impurities level.

A recent set of measurements, not included in this work and yet to be carefully verified , indicate that purer Ne, with impurities or quenchers content of 0.01% has very low reachable gain. We may conclude that some quenching is indispensable for the stable high-gain operation in Ne, preferably provided by controlled admixture of N$_2$ or CH$_4$, at levels of a few percent. Similar effect was observed in Ar, which also emits energetic photons (wavelength ~ 130 nm).

### 3.7 Photoelectron extraction and collection efficiency

One of the applications of CsI coated THGEM could be in RICH detectors, where efficient detection of single photons is required over a large area. It requires high photoelectron extraction efficiency from the photocathode surface (e.g. deposited on of the THGEM's top surface) and their efficient focusing into the THGEM holes (collection efficiency). It is well known that electron extraction into gas is influenced by the electric-field strength on the photocathode surface, affecting backscattering on the gas molecules (see some data in [31]).The collection efficiency depends on the field strength and directions at the holes' vicinity as well as on electron diffusion; both are affected by the voltage difference across the THGEM (effective gain) and the gas composition.

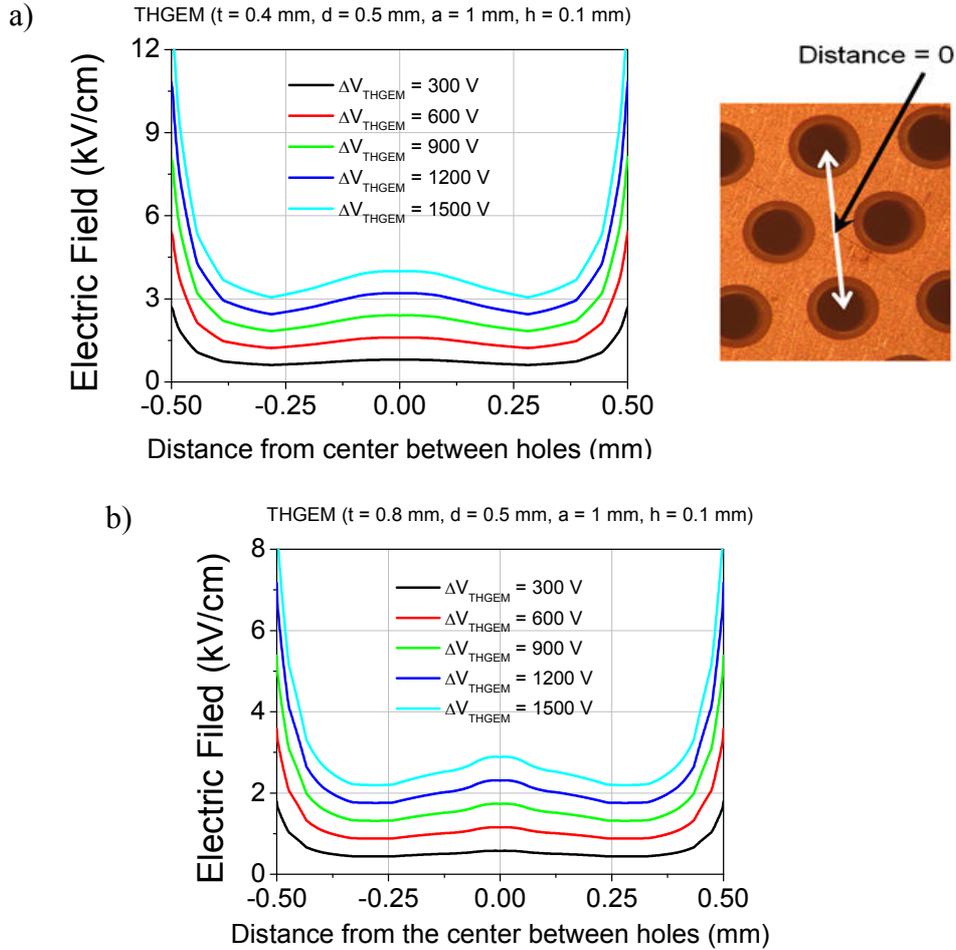

**Figure 17.** Calculated electric field strength on the top-surface of the THGEM along a line between two neighboring holes (scheme on the right); the fields are calculated for different $\Delta V_{THGEM}$ values. Drift and induction fields were 0 kV/cm. a) THGEM parameters: t = 0.4 mm, d = 0.5 mm, a = 1 mm and h = 0.1 mm; b) THGEM parameters: t = 0.8 mm, d = 0.5 mm, a = 1 mm, h = 0.1 mm.

Figures 17 depicts the electric field strength calculated with Maxwell simulation software [30], on the cathode surface of 0.4 or 0.8 mm thick THGEM (hole diameter d = 0.5 mm, pitch a = 1 mm and rim a = 0.1 mm) along a line interconnecting two neighboring holes (see photograph in Fig. 17a). The calculation was made for different $\Delta V_{THGEM}$ values, assuming zero values for the drift and induction fields ($E_{drift}$ = 0 is an optimal value for electron collection, as shown in [4]). As show in the graphs (Fig. 17a), even at very low voltages (300 Volt), the electric field strength at the multiplier's surface is above 0.6 kV/cm for a 0.4 mm thick THGEM and above 0.4 kV/cm for a 0.8 mm thick one; these are important for a good photoelectron extraction from the photocathode. Measurements of extraction- and collection-efficiency were made in the setup shown in Fig. 2.

Figure 18 depicts the photoelectron extraction efficiency ($\varepsilon_{Ex}(V_{drift}) = I_{gas}(V_{drift})/I_{vac}$) from the CsI photocathode as function of the drift field, with no voltage across the THGEM, in Ne, Ne-CH$_4$ mixtures and Ar/CH$_4$ (95:5), at atmospheric pressure, irradiated with a UV Ar(Hg) lamp as shown in Fig. 2a. The extraction in gas was normalized to that in vacuum. For most of the gases, the extraction efficiency showed a sharp rise (up to a field of ~ 0.2 kV/cm) followed by a monotonously growing plateau. In Ne, the extraction efficiency reached a value of the order of 50%, followed by a sharp rise, due to the onset of gas multiplication (e.g. see [32]). The highest extraction efficiencies (~ 75%) were measured in Ne-methane mixtures (5 and 23% CH$_4$) and similarly in Ar/5%CH$_4$ at drift-field values above 0.8 kV/cm. The results obtained in this work are in agreement with those presented in the literature [33].

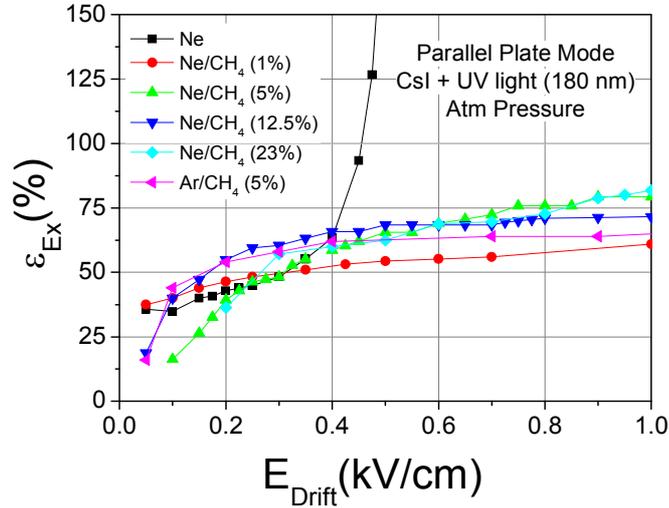

**Figure 18.** Photoelectron extraction efficiency $\varepsilon_{Ex}(V_{drift}) = I_{gas}(V_{drift})/I_{vac}$ from the CsI photocathode (Fig. 2a), as function of the drift field; it was measured in parallel-plate mode (no voltage across the THGEM) in Ne, Ne-CH$_4$ mixtures and in Ar/5%CH$_4$ at atmospheric pressure. The data were normalized to extraction in vacuum.

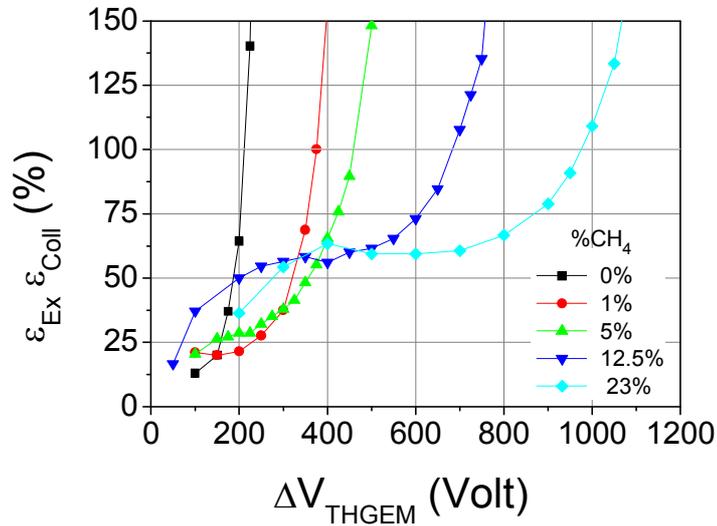

**Figure 19.** Combined photoelectron extraction- and collection-efficiency from a CsI photocathode into THGEM holes $\varepsilon_{Ex}\cdot\varepsilon_{Coll}=I_{THGEM}/I_{vac}$, measured in the setup of Fig. 2b. Data are shown vs $\Delta V_{THGEM}$ in Ne and Ne-CH$_4$ mixtures, at atmospheric pressure. The drift field was kept at 0 V/cm. THGEM parameters: t = 0.8 mm, d = 0.5 mm, a = 1 mm and h = 0.1 mm.

A correct measurement of electron collection efficiency can be done only by the pulse-counting method [4]. But with some small field in the THGEM holes, below multiplication, (gain = 1) the collection efficiency may be evaluated from current measurements, from the relation: $\varepsilon_{Ex} \cdot \varepsilon_{Coll} = I_{THGEM}/I_{vac}$. Fig. 19 shows the normalized THGEM current as function of the THGEM voltage for various Ne-mixtures, in which we assume that the plateau region of the curves represents the current at gain = 1; the sharp rise of the curves reflects the onset of gas multiplication in the THGEM. The measurements were carried out in the setup shown in Fig. 2b, in current mode. Beyond the plateau at gain = 1 we suppose that the extraction and collection efficiencies are further increasing, because $\varepsilon_{Ex}$ increases with the voltage (fig. 18) and because $\varepsilon_{Coll}$ is expected to increase with increasing dipole electric field. Thus one can consider the presented results as a lower limit of the combined extraction and collection yield.

As shown in Fig. 19, the highest combined extraction and collection efficiencies (~ 60%) were measured in Ne-mixtures with 12.5 and 23% of CH$_4$, but at the expense of higher operation voltages. In Ne and with low CH$_4$ concentrations e.g. 1 and 5%, the combined extraction and collection efficiencies were below 25%. Measurements under gain conditions, in pulse-counting mode, [4] are underway in Ne-mixtures.

### 3.8 Localization properties of a Ne-filled detector

The localization properties of a THGEM-based imaging detector (Fig. 3) were characterized in Ne and Ne/5%CH$_4$ at atmospheric pressure, in gas-flow mode. The homogeneity and noise response of the imaging system were obtained from the analysis of an empty-field image, acquired by uniformly irradiating the entire active area of the detector (100x100 mm$^2$) with ca. 9 keV Bremsstrahlung X-rays at a flux rate of ~ 1 s$^{-1}$ mm$^{-2}$. The voltages were set to obtain an operative effective gain of the detector of ~ 10$^5$.

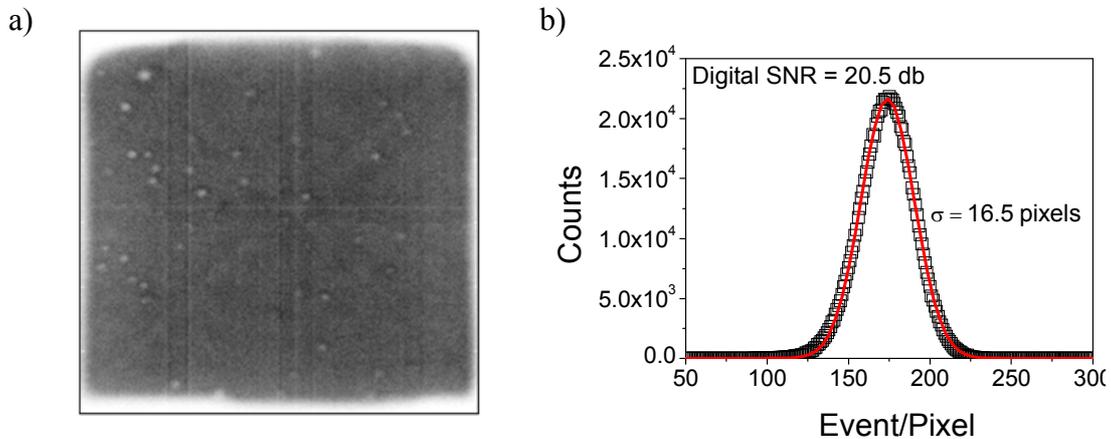

**Figure 20.** a) Grey-level flat-field image of the double-THGEM imaging detector of Fig. 3. b) Histogram of the flat-field image depicted in fig. a); the Gaussian fit of the histogram is shown (red line).

The resulting digital image is presented as a grey-level (8-bit) image, shown in Fig. 20a. The flat-field image shows some bright spots scattered randomly across the active area; they correspond to areas with lower detection efficiency, most probably due to imperfections of the resistive anode [22]. Beside these defects, the flat-field image appears to be homogeneous, reflecting a good homogeneity of the effective gain over the entire detector's area. Fig. 20b depicts the digital histogram of the flat-field image obtained by the analysis of Fig. 20a, perfectly fitted by a Gaussian. (Details of the analysis are given elsewhere [22]). With an average number of counts per single pixel A = 174, and a standard deviation of σ = 16.5, the average gain spread is 9.5%. The characteristic signal-to-noise ratio (SNR) of the imaging system is:

$$SNR = 20 \log_{10}(A/\sigma) = 20.5 \qquad (2)$$

The response of the imaging system, expressed in terms of integral non-linearity (INL), was evaluated from the analysis of a 1 mm thick brass-mask image, with 1 mm holes (see Fig.21a). Fig. 21b depicts the projection along the x-axis of the central region of the image. These data were fitted with seven Gaussian curves, in order to get the centroids of the holes; the obtained centroids were compared to the real hole locations, from which the corresponding INL value was calculated: the average INL value is less than 0.1% over the entire active area of the detector.

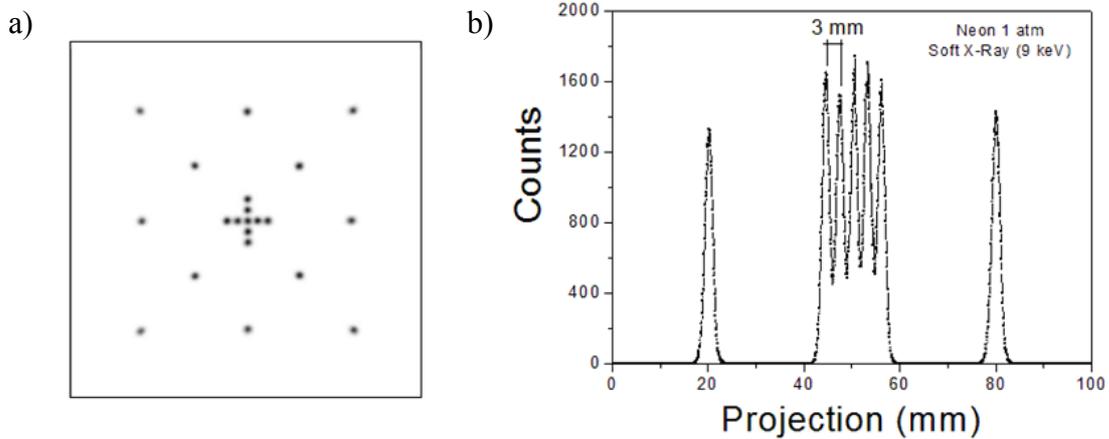

a)     b)

**Figure 21.** Mask image (a) of the THGEM imaging detector of Fig. 3, irradiated with 9 keV X-rays; the 1mm thick mask has 1 mm diameter holes. (b) A projection of the central region of the image.

Though the detector was operating at high gain and with excellent stability, the imaging spatial resolution obtained in Ne was only of 1.4 mm FWHM, which is about two-fold worse compared to the one obtained in the same detector operated with Ar/5%CH₄ [22]. The reason for this is the relatively larger range of the photoelectron in Ne as compared to the one in Ar, as well as larger electron diffusion coefficient. The K-shell photoelectron, emitted by a typically 9 keV photon in Ne, has a practical range of about 2 mm, as compared to 0.6 mm in pure Ar [34]. The readout of our detector measures center of charge-cloud. So, a longer range of the photoelectron, which is emitted isotropically in all directions form the point of interaction of the X-rays photon, directly affects the position resolution.

Moreover, the drift velocity in pure Ne is much lower than in Ar-mixtures, leading to large signal rise-times; the latter affects the position resolution in our encoding system based on delay-line readout [22]. Ne with 5% CH₄ admixture considerably reduced the pulse rise-time, from 75 down to 30 ns (Fig. 9). Irradiating the detector with a continuous Bremsstrahlung X-rays spectrum at electron energy of ca. 4 keV, under same experimental conditions, resulted in an intrinsic spatial resolution of around 0.4 mm, far below the pitch of the THGEM holes (1 mm). As a result, as depicted in fig. 22, details of the hexagonal hole-type structure of the THGEM is emerging from a grey-level flat-field image obtained at low X-rays energies (< 4 keV).

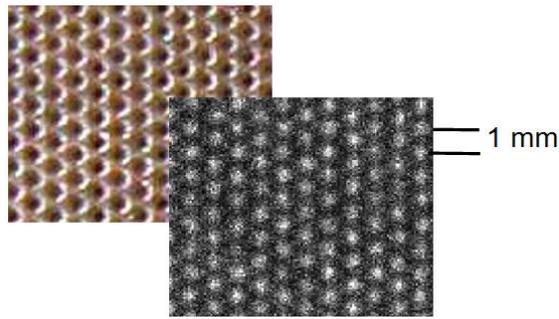

1 mm

**Figure 22.** Details of the hole-type structure appear in the flat field image when irradiating with low energy X-rays (< 4 keV), where photoelectron ranges are of the order of 0.6 mm.

## 4. Conclusions and discussion

The present study included single and double THGEM detectors operated within stainless-steel vessels, evacuated and then continuously flushed with gas. Under these conditions it was estimated that residual air contaminants were present, at levels below 1%; their role in stabilizing the high gain operation was confirmed. Under these conditions, large electron multiplication factors were reached in atmospheric-pressure Ne and Ne-CH$_4$ gas mixtures, in both single- and double-THGEM detectors, at low operation voltages; the detectors operated in a stable way with a ten-fold higher dynamic range (soft X-rays vs. single electrons) and better gain stability compared to Ar/5%CH$_4$. THGEM voltages of ~ 500 and ~ 400 Volt resulted in respective single-photoelectron effective gains in the range of $10^5$ and $10^7$ with single- and double-THGEM (0.4 mm thick) multipliers. 0.8 mm thick single-THGEMs provided single-photoelectron effective gains of the order of $10^7$ in Ne and Ne-methane mixtures. It was shown that the avalanche extension out of the holes, of increasing size with increasing hole diameter, induces secondary effects in Ne; we suppose these are due to photon-feedback, as avalanche extension is accompanied by photons produced out of the holes and possibly hitting the PC surface. An interesting observation was made, however, that the secondary effects from a CsI photocathode was not considerably stronger compared to those from a Cu cathode (see Fig. 12a). This may be explained either by the fact that CsI and Cu have comparable QE values in the Ne far-UV (74-85 nm) emission wavelength, or by the actual quenching of the Ne far-UV emission by gas impurities [33]; more experiments are needed to clarify the photon spectrum and their role in Ne-based THGEM detectors; it is important to note that the feedback is efficiently quenched with small admixtures of N$_2$ or CH$_4$.

Ion-induced secondary effects may also play a role in Ne based detectors: Ne ions have high potential energy and could in principle induce secondary electron emission via Auger neutralization process on the CsI PC. This process was estimated to have rather low efficiency, of about 1% and therefore can be seen only at relatively high gains (above $10^3$) [35]. As was mentioned in paragraph 2, gas contaminant might be responsible for the suppression of such ion-induced feedback, by charge exchange process. It is our general conclusion that impurities such as residual water and other air-contaminants have a beneficial effect on the operation stability, total gain and the energy resolution, although it is hard to control and optimize their content in the gas.

An interesting possibility to increase the gain and reduce the operation voltages of Ne-filled THGEM detectors might be the use of Ne-H$_2$ mixtures, known to have a Penning effect (e.g. with 0.1%-0.2% H$_2$), as demonstrated with a 3-GEM detector at low temperatures and at high pressure [19]. The larger dynamic range in Ne-mixtures, due to broader electron-avalanche spread (probably resulting from a different Raether limit), could have important implications on the operation stability in applications like single-photon imaging (e.g. in RICH) or particle tracking – at the presence of highly ionizing background.

Photoelectron extraction efficiency values in Ne-CH$_4$ showed to be rather similar to those in Ar-CH$_4$ mixtures, of the order of 75% for realistic fields at the photocathode surface of ~ 1kV/cm. Preliminary measurements of combined photoelectron extraction and collection efficiencies pointed out at "lower-limit" numbers (at THGEM gain = 1) ranging between 26% to 60%. More complex rigorous investigations of electron collection under high gain, in pulse-counting mode, are in course. We speculate that we may benefit from the fact that multiplication onset in Ne and Ne-CH$_4$ mixtures occurs at fields of the same order or lower than the field at the hole's vicinity. This may cause "pre-amplification" of the extracted photoelectrons on their way to the holes – resulting in increased detection efficiencies of single photons. The same phenomenon was previously found to be successful in enhancing single electron detection efficiency in GEMs [36].

In summary, Ne and Ne with some admixtures could be good candidates for THGEM-based detectors in many applications. For example, such devices are being investigated for UV detection in RICH (by members of the CERN-RD51 collaboration) and for neutron imaging (Weizmann/PTB/Milano). They have potential applications in detectors for rare events, including sensing elements in noble-liquid detectors. For instance, successful THGEM operation at cryogenic conditions was recently demonstrated [7,8].

## Acknowledgments

This work was partially supported by the Israel Science Foundation, grant Nº 402/05, by grant No 3-3418 of the Israel Ministry of Science, Culture & Sport within a France-Israel Scientific Cooperation, by the MINERVA Foundation, grant Nº8566, by the Benoziyo Center for High Energy Research and by projects POCI/FP/81980/07 and CERN/FP/83645/08 through FCT and FEDER programs. M. Cortesi and G. Bartesaghi warmly acknowledge the Fellowship of the Lombroso Foundation. A. Breskin is the W.P. Reuther Professor of Research in the peaceful use of Atomic Energy.

## References

[1] R. Chechik, A. Breskin, C. Shalem and D. Mörmann, Thick GEM-like hole multipliers: properties and possible applications, *NIMA* **535** (2004) 303-305.

[2] L. Periale, V. Peskov, P. Carlson, T. Francke, P. Pavlopoulos, P. Picchi and F. Pietropaolo, Detection of the primary scintillation light from dense Ar, Kr and Xe with novel photosensitive gaseous detectors, *NIMA* **478** (2002) 377-383.

[3] P. Jeanneret, Time projection chambers and detection of neutrinos, Ph.D. Thesis, Neuchâtel University, 2001.

[4] C. Shalem, R. Chechik, A. Breskin and K. Michaeli, Advances in Thick GEM-like gaseous electron multipliers - Part I: atmospheric pressure operation, *NIMA* **558** (2006) 475-489.

[5] R. Alon, J. Miyamoto, M. Cortesi, A. Breskin, R. Chechik, I. Carne, J.M. Maia, J.M.F. dos Santos, M. Gai, D. McKinsey and V. Dangendorf, Operation of a Thick Gas Electron Multiplier (THGEM) in Ar, Xe and Ar-Xe, 2008 *JINST* **3** P01005.

[6] A. Breskin, R. Alon, M. Cortesi, R. Chechik, J. Miyamoto, V. Dangendorf, J.M. Maia and J.M.F. Dos Santos, A concise review on THGEM detectors, *NIMA* **598** (2009) 107–111.

[7] L. Periale, V. Peskov, C. Iacobaeus, B. Lund-Jensen, P. Pavlopoulos, P. Picchi and F. Pietropaolo, A study of the operation of especially designed photosensitive gaseous detectors at cryogenic temperatures, *NIMA* **567** (2006) 381-385.

[8] A. Bondar, A. Buzulutskov, A. Grebenuk, D. Pavlyuchenko, Y. Tikhonov and A. Breskin, Thick GEM versus thin GEM in two-phase argon avalanche detectors, 2008 *JINST 3* P07001.

[9] R. Chechik, A. Breskin and C. Shalem, Thick GEM-like multipliers - a simple solution for large area UV-RICH detectors, *NIMA* **553** (2005) 35-40.

[10] F. Tessarotto, talk at 5th International Conference on New Developments In Photodetection, Aix-les-Bains, France, June 15-20, 2008.

[11] A. Di Mauro, talk at the ALICE Upgrades Kickoff meeting, CERN, Geneva, Switzerland, March 24, 2009 (http://indico.cern.ch/materialDisplay.py?contribId=5&materialId=slides&confId=53025).

[12] A. White, Digital Hadron Calorimetry using Gas Electron Multiplier Technology, CALICE Collaboration meeting, NIU, DeKalb, Illinois, 2005. (http://nicadd.niu.edu/calice/)

[13] J.T. White , J. Gao, J. Maxin, J. Miller, G. Salinas and H. Wang, SIGN a WIMP Detector Based on High Pressure Gaseous Neon, Proceeding of the International Conference DARK2004, College Station, USA, Oct 3-9, 2004.

[14] R. Galea, J. Dodd, W. Willis, P. Rehak and V. Tcherniatine, Light Yield Measurements of GEM Avalanches at Cryogenic Temperatures and High Densities in Neon Based Gas Mixtures, *IEEE Nuclear Science Symposium Conference Record* **1** (2007) 239-241.

[15] M. Gai, D. N. McKinsey, K. Ni, D. A. R. Rubin, T. Wongjirad, R. Alon, A. Breskin, M. Cortesi and J. Miyamoto, Toward application of a Thick Gas Electron Multiplier (THGEM) readout for a Dark Matter detector, *Proceedings of the 23rd Winter Workshop on Nuclear Dynamics*, Big Sky, Montana (USA), February 11-18, 2007 (arXiv:0706.1106v1).

[16] R. Oliveira, V. Peskov, F. Pietropaolo and P. Picchi, First tests of thick GEMs with electrodes made of a resistive kapton, *NIMA* **576** (2007) 362-366.

[17] R. Bellazzini, F. Angelini, L. Baldini, A. Brez, E. Costa, L. Latronico, N. Lumb, M.M. Massai, N. Omodei, P. Soffitta, and G. Spandre, X-ray polarimetry with a Micro Pattern Gas Detector with pixel readout, *IEEE Trans. Nucl. Scien.* **49** (2002) 1216-1220.

[18] H. Sakurai, F. Tokanai, S. Gunji, T. Sumiyoshi, Y. Fujita, T. Okada, H. Sugiyama, Y. Ohishi and T. Atsumi, Gain of a gas photomultiplier with CsI photocathode, *J. of Phys.: Conference Series* **65** (2007) 012020.

[19] A. Buzulutskov, J. Dodd, R. Galea, Y. Ju, M. Leltchouk, P. Rehak, V. Tcherniatine, W. J. Willis, A. Bondar, D. Pavlyuchenko, R. Snopkov and Y. Tikhonov, "GEM operation in helium and neon at low temperatures, *NIMA* **548** (2005) 487-500.

[20] E.W. MacDaniel, Collision phenomena in ionized gases, Wiley and Sons, New York (1964).

[21] T. Shima, K. Watanabe, T. Irie, H. Sato and Y. Nagai, A gas scintillation drift chamber for the 14N(n, p)14C measurement, *NIMA* **356** (1995) 347-355.

[22] M. Cortesi, R. Alon, R. Chechik, A. Breskin, D. Vartsky, V. Dangendorf, Investigations of a THGEM-based imaging detector, 2007 *JINST 2* P09002.

[23] see (http://www.roentdek.com/)

[24] V. Peskov, P. Fonte, M. Danielsson, C. Iacobaeus, J. Ostling, and M. Wallmark, The study and optimization of new micropattern gaseous detectors for high-rate applications, *IEEE Trans. Nucl. Scien.* **48** (2001) 1070 – 1074.

[25] R. Chechik, M. Cortesi, A. Breskin, D. Vartsky, D. Bar and V. Dangendorf, Thick GEM-like (THGEM) detectors and their possible applications, *Proceedings of the SNIC Symposium 2006*, Stanford, California (USA), 3-6 April 2006, (arXiv:physics/0606162).

[26] P. Fonte, V. Peskov and B.D. Ramsey, The fundamental limitations of high-rate gaseous detectors, *IEEE Trans. Nucl. Scien.* **46** (1999) 321 – 325.

[27] P. Fonte, V. Peskov and B.D. Ramsey, Rate and gain limitations of MSGCs and MGCs combined with GEM and other preamplification structures, *NIMA* **419** (1998) 405-408.

[28] Y. Ivaniouchenkov, P. Fonte, V. Peskov and B.D. Ramsey, Breakdown limit studies in high-rate gaseous detectors, *NIMA* **422** (1999) 300-3004.

[29] V. Peskov, B.D. Ramsey and P. Fonte, Breakdown features of various microstrip-type gas counter designs and their improvements *IEEE Trans. Nucl. Scien.* **45** (1998) 244 – 248.

[30] MAXWELL 3D, ANSOFT Co. Pittsburgh, PA, USA.

[31] D. Mörmann, A. Breskin, R. Chechik and C. Shalem, Operation principles and properties of the multi-GEM gaseous photomultiplier with reflective photocathode, *NIMA* **530** (2004) 258-274.

[32] L.C.C. Coelho, H.M.N.B.L. Ferreira, J.A.M. Lopes, T.H.V.T. Dias, L.F.R. Ferreira, J.M.F. dos Santos, A. Breskin, R. Chechik, Measurement of the photoelectron-collection efficiency in noble gases and methane, *NIMA* **581** (2007) 190-193.

[33] V. Peskov, G. Charpak, W. Dominik and F. Sauli, Investigation of light emission from a parallel-plate avalanche chamber filled with noble gases and with TEA, TMAE, and $H_2O$ vapours at atmospheric pressure, *NIMA* **277** (1989) 547-556.

[34] T. Ferbel: Experimental Techniques in High Energy Physics, Addison-Wesley Pub. Comp., Inc., Menlo Park, California (1987).

[35] A. Lyashenko, A. Breskin, R. Chechik and T. H. V. T. Dias, Ion-induced secondary electron emission from K-Cs-Sb, Na-K-Sb and Cs-Sb photocathodes and its relevance to the operation of gaseous avalanche photomultipliers, submitted to *J. Appl. Phys.* (arXiv:0904.4881v1)

[36] C. Richter, A. Breskin, R. Chechik, D. Mörmann, G. Garty and A. Sharma, On the efficient electron transfer through GEM, *NIMA* **478** (2002) 538-558.